\documentclass[useAMS,usenatbib]{mn2e}
\usepackage{epsfig} 
\usepackage{graphicx}
\usepackage{color} 
\usepackage{amsmath} 
\def\eg{{\it e.g.,}~}
\def\etal{{\em et al.}~}
\def\ie{{\em i.e.,}~}
\def\lsim{\lower.5ex\hbox{$\; \buildrel < \over \sim \;$}}
\def\gsim{\lower.5ex\hbox{$\; \buildrel > \over \sim \;$}}
\def \simeq{\lower.3ex\hbox{$\; \buildrel \sim \over - \;$}}

\def\araa{{ARA\&A}}%
\def\apj{{ApJ}}%
\def\aap{{A\&A}}%
\def\mnras{{MNRAS}}%
\def\na{{New A}}%
\def\pasj{{PASJ}}%
\title[Radiatively and thermally driven jets]
{Radiatively and thermally driven self-consistent bipolar outflows from accretion discs around compact objects}
\author[Kumar, R. and Chattopadhyay, I., Mandal, S.]{Rajiv Kumar$^{1}$\thanks{E-mail: rajiv.k@aries.res.in (RK);
indra@aries.res.in (IC); samir@iist.ac.in (SM)},
Indranil Chattopadhyay$^1$, Samir Mandal$^2$\\
$^{1}$Aryabhatta Research Institute of Observational Sciences (ARIES),
Manora Peak, Nainital $-$ 263002, India\\
$^{2}$Indian Institute of Space Science $\&$ Technology (IIST), Trivandrum, India.
 }
\begin{document}
\date{Accepted -----. Received ------; in original form ----}

\pagerange{\pageref{firstpage}--\pageref{lastpage}} \pubyear{2012}

\maketitle

\label{firstpage}

\begin{abstract}
We investigate the role of radiative driving of shock ejected bipolar outflows from advective accretion discs
in a self consistent manner.
Radiations from the inner disc affects the subsonic part of the jet while those from the pre-shock disc
affects the supersonic part, and there by constitutes a multi stage acceleration process.
We show that the radiation from the inner disc not only accelerate but also increase the mass outflow rate,
while the radiation from the pre-shock disc only increases the kinetic energy of the flow. With proper 
proportions of these two radiations, very high terminal speed is possible. We also estimated the post-shock
luminosity from the pre-shock radiations, and showed that with the increase of viscosity parameter
the disc becomes more luminous, and the resulting jet simultaneously becomes
faster. This mimics the production of steady mildly relativistic but stronger jets as micro-quasars
moves from low hard to intermediate hard spectral states.
\end{abstract}

%% each reference.
%%%%%%%%%%%%%%%%%%%SECTION 1 %%%%%%%%%%%%%%%%%%
\section{Introduction}

Jets are ubiquitous and are observed to accompany a variety of astrophysical objects 
such as, stellar-mass and super-massive black hole candidates, neutron stars, white dwarfs,
young stellar objects (YSOs) etc.
These outflows exhibit 
different physical scales and power 
\eg at
one extreme, AGNs have jets with typical sizes $\sim$ few Kpc to few $\times ~ 100$ Kpc, jet velocities
comparable to
$c$ (where $c$ is the light speed),
luminosity range $10^{43-48}$erg/s and central mass in the range $10^{6-9}M_{\odot}$ (where $M_{\odot}$ is the
mass of Sun), while in the other extreme, YSO jets have typical size $\le 1$pc, outflow velocity $<10^{-3}c$,
luminosity range $(0.1-2)\times 10^{36}$erg/s, and emerge from protostars with mass $\sim 1M_{\odot}$. In other words,
only jets around compact objects are truly relativistic. 
Jets around black hole candidates, be it from AGNs or micro-quasars, are shrouded in mystery. Since black holes
do not have hard surface and/or any atmosphere, therefore jets can only originate from the matter accreting onto it.
Moreover, the terminal speed of jets from around compact objects, though relativistic, but can vary widely too. For
\eg
jets around GRS 1915+105 or M87 exhibit terminal speed above ninety percent the speed of light \citep{mr94,b93},
while the jet around SS433 is merely around $0.26c$ \citep{m84}.
Therefore, not only there is no consensus about the origin of jets,
even the acceleration mechanism of jets are not well understood.

There are few interesting properties of jets around compact objects. \citet{jbl99} showed that the jet around
M87 seems to originate within a region less than $100~r_s$ ($r_s=2GM/c^2$ is the Schwarzschild radius) around the
central object. In other words, entire accretion disc do not participate in generation of jets, but only the inner 
region of the accretion disc participate in jet generation. Although the connection between jet states and
spectral states of the accretion disc has not been conclusively established for massive black hole candidates like the AGNs,
but for micro-quasars, this connection has firmly been established \citep{gfp03,fgr10,rsfp10}. Persistent,
quasi steady, mildly
relativistic jets are observed in `hard spectral state' (maximum power in the hard power law tail)
of the accretion disc. The jet seems to get stronger as the spectral state of accretion disc moves to the
intermediate states. And truly relativistic jet blobs are generated during state transition to the steep power law
state. No jet activity is observed in the canonical soft state (maximum power in modified black body component)
\citep{gfp03,fb04,rsfp10}. Such close corelation of jet with the radiative states of the disc, points to the fact that
the accretion disc physics is responsible for jet generation.

Matter accreting onto a compact object should possess some angular momentum and due to differential rotation
some form of anomalous viscosity too. The first viscous disc model seriously considered by the community
is the standard thin disc \citep{ss73,nt73}, although thin disc's inability to explain the origin of
power-law photons,
as well as, the theoretical inconsistencies like adhoc inner boundary condition, remained an overbearing concern.
It was understood that a source of hot electron distribution \ie a Comptonizing corona, is required to explain the hard power law photons in the spectra
of black hole candidates \citep{st80}.
Since the
boundary condition of black hole accretion is necessarily transonic, disc models with significant advective term gained
popularity. The most popular model among the advective disc models was known as ADAF or advection dominated
accretion flow \citep{nkh97}. This model is characterized by a single sonic point close to the  horizon and
subsonic elsewhere. However, it has been shown earlier that multiple sonic point may exist for inviscid
rotating flow \citep{lt80}, and was confirmed that this is also true for dissipative advective accretion discs
\citep{c96,lgy99,cd04}. And therefore it was shown by various authors that ADAF is only a subset of
general advective solution \citep{lgy99,dbl09,kc13}. General advective solutions which admits multiple sonic points
may harbour standing or time-dependent shock solutions \citep{f87,c89,c96,ft04,letal08,c08,ny09,cc11}.
Shock in accretion has some
interesting consequences, for \eg the post shock hot electrons may inverse Comptonize soft photons to produce
the non-thermal hard radiation tail \citep{ct95,cm06}.
A dominant shock associates with a low supply of soft photons from external Keplerian disc,
and the hot post-shock region inverse-Comptonizes the intercepted
photons and produces the low/hard state. And an increased supply of Keplerian matter means supply of extra
soft photons to cool down the post-shock region which results in weakening of, or,
complete removal of the shock to produce the canonical high/soft state.
In fact, the hardness intensity diagram (HID) of GRO J 1655-40 was well explained by shocked accretion disc
\citep{mc10}.
These studies also showed that the steady state shock solutions
are possible in a limited
range of the parameter space, while oscillating or time dependent shocks are possible for a wide range of 
parameters or boundary conditions \citep{msc96,nrks05,ny09}. And since these post-shock region also emits
high energy radiations, therefore quasi-periodic oscillation of the shocked disc will give rise to 
quasi-periodic oscillation in hard radiations as well, and was furthered as a model
for quasi-periodic oscillations or QPOs \citep{msc96,cm00,nrks05,ny09}.  
It was shown from observations that the spectral index increases (hard to soft transition) with the
increase of QPO frequency, but remarkably the spectral index saturates (indicating the attainment
of high/soft state) and then there are no QPO detected \citep{st09}. The evolution of the QPO frequency with the
spectral state for outburst sources like XTEJ1550-654, GRO 1655-40 etc, has been explained well with a model
of inward drift of oscillating shock which translates into a spectral state transition from low/hard to
intermediate hard. And the final disappearance of QPO with the disappearance of the inner disc or
the post-shock disc \citep{cdnp08,cdp09}. 
And it has been shown that indeed for the same outer boundary
condition, the shock drifts inwards with the increase of the viscosity parameter \citep{cd07,kc13}.
This implies that, with the increase of viscosity the pre-shock disc size increases,
which increases the supply of soft photons.
As the post-shock disc gets reduced, eventually the total output of the thermal Comptonization
(dominant spectra in low/hard state) decreases too. And since according to the advective accretion disc model,
oscillation of post-shock disc gives rise to QPO, so the decrease of shock location increases the QPO frequency,
until the shock itself disappears.
Hence the transition from hard to soft state, or in other words, the increase of spectral index should be
correlated with the increase of QPO frequency, and finally the spectral index saturates \citep{tf04,st07}
as the contribution to the radiation
from post-shock
disc becomes negligible \citep{cdnp08}.

In our previous paper \citep{kc13}, we have presented all possible advective viscous accretion solutions,
including shocked and shock free solutions.
A shocked accretion disc is more interesting, because the post-shock disc being hot can drive bipolar-outflows
\citep{mrc96,c99,cd07,dc08,kc13}. 
Since shock forms typically at $x_s \lsim$few $\times 10 r_s$, 
so the assertion that jet base is the inner part of the disc,
favours the observational evidence from M87 that jet base is indeed $< 100r_s$ \citep{jbl99,detal12}. 
Since it has also been shown that
AGNs are just a scaled up version of the galactic black holes \citep{mkkuf06}, therefore, following
the above evidence one may argue that even for galactic black holes the jet base should be close to the central
object.
However, it is not just theoretical expectation that the inner region around AGNs and micro-quasars should be
similar, therefore, the jet base for microquasars too would be close to the horizon.
Even in case of galactic Black Hole Candidates (BHC), the detection of strong radio flares (read jets) 
with simultaneous disappearance of QPOs and the absence of the Comptonized
component, points out to the fact that the same region which gave rise to the Comptonized component
and generated the QPO has been ejected as relativistic jets \citep{fmpctb99,vrnc01}.  The shocked accretion disc model
seems to satisfy all these criteria, starting with the direct evidence of M87 jet originating close to the black hole,
the connection between growth of QPO frequency, Comptonized component and the radio flare, to the disappearance
of jet activity in the soft state (no or very weak shock). \citet{dcnc01} showed (Fig. 5 of their paper)
that the correlation between
thermally driven jets from shocked accretion discs and the spectral states from the same discs seems to follow 
the conclusions of observations of 10 microquasars \citep{gfp03}.
Since post-shock disc produces the jets therefore when shock is absent or very weak,
jets disappear mimicking the soft state.
But thermally driven jets can achieve terminal speed up to $v_{\infty}\lsim [2\{a^2_b/(\gamma -1)-\Phi_b\}]^{1/2}$,
where
$a_b$ is the sound speed at the base, which for a shocked disc is the sound speed of the post-shock disc,
and $\Phi_b$ is the gravitational potential at the jet base.
Now the maximum sound speed physically possible is $a_{\rm max}=c/{\sqrt{3}}$, and that the value of
$\Phi_b$ is quite significant because
the jet base is rather
close to the horizon \citep{jbl99}. Given these facts, the expression of terminal speed given above indicates that,
truly relativistic jet terminal speed is not possible by only
thermal driving. Furthermore, jet states are correlated with the hard spectral states of the accretion disc,
therefore, can the disc radiation accelerate the outflowing jet material? In this respect, one may
raise the issue that jet activities are not always observed during the state transition, which might be due to the
lack of availability of simultaneous X-ray and radio/infra-red measurements. However, \citet{gfp03} studied
about 10 sources to draw the correlation between jet activity with the hard states, and the conclusion
that jets are not seen during the canonical soft states. We would therefore like to study the
interaction of radiations from the hard/hard intermediate states with the emanating jet, or in our parlance,
the acceleration of jets with the radiation from a shocked disc, and how this radiative acceleration of jets is
influenced with the change in viscosity parameter of the disc.
It is to be remembered that, since we are investigating the disc in
steady state, we are actually studying the generation and acceleration of steady, mildly relativistic
jets generally associated with hard to hard intermediate states. The strong radio flares associated
with relativistic blob ejections during hard intermediate to soft intermediate state transition \citep{mil12} is intrinsically
a time dependent phenomenon and has not been addressed in this paper. Since we have not incorporated
all the physical aspects to mimic the transient phenomena, in this limited sense the results depend on the
model assumptions.

Interaction of the disc radiation with the outflowing matter has been followed by many authors. \citet{i80}
studied radiative acceleration of jets above a Keplerian disc, but in absence of radiation drag.
\citet{sw81}
showed that radiation drag is important for jets powered by radiation from a thick accretion disc \citep{lb78}.
Including radiation momentum deposition on an axial particle jet illuminated by an infinite thin disc in presence of
drag term, \citet{i89}
showed that the upper limit of the terminal speed is around $45\%$ of $c$. 
Investigations on radiatively driven jets were extended by \citet{f96,fth01} to generate relativistic terminal speeds
for particle jets. Moreover, the interaction of radiations from shocked accretion disc and jets
were studied too \citep{cc00,cc02,cc03,cdc04,c05}. Since shocked discs have two radiation sources, namely,
the post-shock and pre-shock disc, therefore the redistribution of radiative power amongst these sources,
could efficiently accelerate the jet around its axis.  However, jets studied in the above
mentioned papers were decoupled
from the disc. In this paper we would like to study radiatively driven jet which is generated self consistently
from the underlying accretion disc. In other words, we combine the accretion-ejection
physics described by \citet{kc13} with the effect
of radiative momentum deposition on the jets as shown by \citet{cc02}.

In the next section, we present the simplifying assumptions and equations of motion. In section 3,
we present the methodology of solution. And in the last section we present the solutions
and discussion.

%%%%%%%%%%%%%%%%SECTION 2 %%%%%%%%%%%%%%%%%%%%%%%%%%%%%%%%%%%%%%%%%%%%%%

\section{Equations of motion and Assumptions}

In this paper, we have assumed the axis-symmetric disc-jet system to be in steady state.
The black hole is assumed to be non-rotating described by the pseudo-Newtonian
potential introduced by \citet{pw80}. 
The viscosity prescription in the accretion disc is described by the Shakura-Sunyaev $\alpha$ viscosity.
The jets are tenuous and should have less differential rotation than the accretion disc,
as a result the viscosity in jets can be ignored. To negate any resulting torque, the
angular momentum at the jet base is assumed same as that of the local value of angular
momentum of the disc.
We estimate the radiative moments,
namely, the radiative energy density, flux, and radiative pressure from the disc.
The jets being optically thin are subjected to radiation force including the drag terms.
Figure 1, shows the schematic diagram of the disc-jet system, where the pre-shock, post-shock disc and the jet
are clearly marked. The black hole is at BH. The accretion disc occupies the space on or about the equatorial plane,
and the jet
flow geometry is about the axis of symmetry. 
We have used the geometric unit system where, $2G=M=c=1$ ($M$ is the mass of the black
hole, $G$ is the Gravitational constant). Therefore, in this representation the unit of length
is, the Schwarzschild radius or $r_s=2GM/c^2$, the unit of
mass is that of the central black hole $M$, and that of time is
$r_s/c=2GM/c^3$, respectively, consequently the unit of speed is $c$.

\begin{figure}
\begin{center}
\epsfig{figure=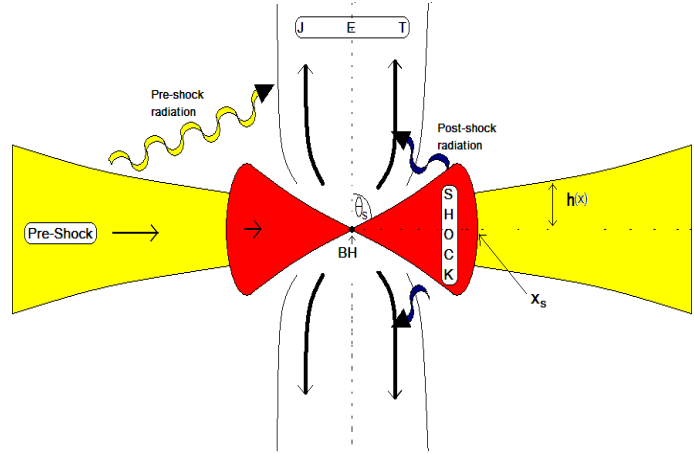,height=6.cm,width=9.cm,angle=0}
\end{center}
\caption{Cartoon for disc-jet system. BH is the acronym for black hole. Both pre-shock and post-shock
 disc shines radiation on the bipolar jet.The shock location $x_s$, the axis (vertical dotted) and equatorial plane
(horizontal dotted) are shown. The half height $h(x)$, and the opening half angle $\theta_s$ of the funnel like
region between the inner torus is shown.}
\label{lab:fig1}
\end{figure}

\subsection{Equations of motion for accretion and ejection system} 
\label{subsec:acceq}

Since we are investigating self-consistent accretion-ejection system, we should present the 
governing equations together. However, since the accretion and jets follow different 
flow geometries, we present them separately.
In section 2.1.1, we present equations of motions, which when solved gives us the
accretion solution. In section 2.1.2, the equations governing the jet solution are presented.
And finally in section 2.2, we present the solution procedure, \ie how a jet solution is simultaneously obtained
from accretion disc through shocks.

\subsubsection{Equations governing the accretion disc}
\label{subsubsec:eqacc}
The equations of motion of the accreting matter around the equatorial plane,
in cylindrical coordinates ($x$, $\phi$, $z$), have been extensively presented
in the literature \citep{c96,kc13}. Here we give a brief account of them.

These equations are the radial momentum equation, the accretion rate equation for a disc in vertical equilibrium,
the angular momentum
conservation equation and the entropy generation equation, and they are
%\begin{center}
\begin{equation}
u\frac{du}{dx} + \frac{1}{\rho}\frac{dp}{dx} + \frac{1}{2(x-1)^2}-\frac{\lambda^{2}(x)}{x^{3}} = 0,
\label{rme.eq}
\end{equation}
\begin{equation}
\dot{M}=2\pi\Sigma u x,
\label{mf.eq}
\end{equation}

\begin{equation}
u\frac{d\lambda(x)}{dx}+\frac{1}{\Sigma x}\frac{d(x^{2}W_{x\phi})}{dx}=0
\label{amde.eq}
\end{equation}

\begin{equation}
\Sigma u T\frac{ds}{dx}=Q^{+}-Q^{-}.
\label{ege.eq}
\end{equation}
The local variables $u,~p,~\rho$ and $\lambda$ in the above equations are the radial velocity,
isotropic pressure, density and specific angular momentum of the flow, respectively.
Here, $\Sigma=2\rho h$ is vertically integrated density with $h$ being the local disc height 
from equatorial plane and $W_{x\phi}$ is the viscous stress tensor.
Moreover, $s$, $T$ are the entropy density and the local
temperature respectively. The local heat gained (due to viscosity) and
lost by the flow are given by $Q^{+}=W_{x\phi}^{2}/{\eta}$ and $Q^{-}$.
The viscous stress, the local half height and the sound speed are given by
\begin{equation}
W_{x \phi}= \eta x\frac{d\Omega}{dx},~~h(x)=\sqrt{\frac{2}{\gamma}}ax^{1/2}(x-1), ~~a=\sqrt{\frac{\gamma p}{\rho}}
\label{wha.eq}
\end{equation}
where, 
$\eta=\rho\nu h$ is the dynamic viscosity coefficient, 
$\nu=\alpha a^{2}/(\gamma \Omega_{k})$ is the kinematic viscosity, $\alpha$ is the
Shakura-Sunyaev viscosity parameter, $\Omega$, $\Omega_{k}$ and $\gamma$ are
the local angular velocity, local Keplerian angular velocity and adiabatic 
index, respectively. Here, all the variables have been made dimensionless by employing the unit system
mentioned above section 2.1, \eg if $\tilde{r}$, $\tilde{u}$ etc are dimensional radial coordinate and velocity, then 
$x=\tilde{r}/r_s$, $u=\tilde{u}/c$ etc.

Integrating equation (\ref{rme.eq}) with the help of  Eq. (\ref{mf.eq}-\ref{ege.eq}),
we find another constant of motion called the specific grand energy \citep{gl04} of the flow and is given by
\begin{equation}
E =\frac{u^{2}}{2} + \frac{a^{2}}{\gamma-1}-\frac{\lambda^{2}}{2x^{2}}+\frac{\lambda_{0}\lambda}{x^{2}}
-\frac{0.5}{(x-1)}, 
\label{ge.eq}
\end{equation}
where, a, $\lambda_0$ are local sound speed and the specific angular 
momentum of the flow on the horizon which is one of the constants
of integration.

Using the adiabatic law, the definition of sound speed, and the mass-accretion
equation (Eq. \ref{mf.eq}), entropy-accretion rate is defined
\begin{equation}
 \dot{\cal{M}}(x)=a^{(2n+1)}ux^{3/2}(x-1),
\end{equation}
where $n=1/(\gamma-1)$. ${\dot {\cal M}}$ is the measure of local entropy and is constant for
an inviscid and adiabatic flow.

We rearrange and simplify Eqs. (\ref{rme.eq}-\ref{ege.eq}) to
the following forms (see, Kumar \& Chattopadhyay 2013 for details),
\begin{equation}
 \frac{du}{dx}=f_1(u,a,\lambda,\lambda_0,\alpha,\gamma,x); ~~
\frac{d\lambda}{dx}=f_2(u,a,\lambda,\lambda_0,\alpha,\gamma,x); ~~
\frac{da}{dx}=f_3(u,a,\lambda,\lambda_0,\alpha,\gamma,x)
\label{grad.eq}
\end{equation}

Since accreting matter at $x\rightarrow \infty$ is subsonic, and close to the horizon it is supersonic,
at some point in between, the flow would become transonic, where $du/dx\rightarrow 0/0$.
This gives the so-called sonic point conditions or critical point conditions. And at
$x=x_c$  or sonic point, the gradient is calculated by employing L'Hospital rule.
The accretion solutions are obtained by solving Eqs. (\ref{grad.eq}) with the help of sonic point
conditions.

\subsubsection{Jet equations of motion}
\label{subsubsec:eqjet}
The jet streamline is approximated from the numerical simulations \citep{mrc96}, where the jet flows through
the funnel wall (FW) and the centrifugal barrier (CB). 
A detail jet geometry has been described in \citet{kc13}.
All the jet variables in the equations below have been
made dimensionless by dividing the distances with $r_s$, velocities with $c$, mass with $M$ etc.

The disc photons accelerate the jet by depositing
the radiation momentum onto the jet.
Since the jet is optically thin so the radiation actually penetrates the jet.
 So every fluid parcel of the jet
is accelerated by the radiation from behind, but the radiation field ahead also drags and decelerate
the parcel. The momentum balance equation of the jet, correct up to first order in velocity, is given by
\citep{mm84,cc02}   
\begin{equation}
 v_j\frac{dv_j}{dr}+\frac{1}{\rho_j}\frac{dp_j}{dr}-\frac{\lambda_j^2}{x_j^3}\frac{dx_j}{dr}
+\frac{1}{2(r_j-1)^2}\frac{dr_j}{dr}={\cal{F}}_{r_j}-v_j({\cal{E}}_{r_j}+{\cal{P}}_{r_j})
\label{radmjet.eq}
\end{equation}
Here, ${\cal{F}}_{r_j}=\sigma_{T}F_{r_j}/m_ec, {\cal{E}}_{r_j}=\sigma_TE_{r_j}/m_e$, and ${\cal{P}}_{r_j}
=\sigma_TP_{r_j}/m_e$, where, $\sigma_T$ is the Thompson scattering cross section, $m_e$ is the electron
mass, and $F_{r_j}$, $E_{r_j}$, $P_{r_j}$ are radiative flux, radiative energy density
and radiative pressure, respectively. 
The streamline is computed as $r_{j}=\sqrt{x_{j}^{2}+y_{j}^{2}}$.
The first term in r. h. s of Eq. (\ref{radmjet.eq}) is the acceleration
term, while negative terms are the radiative drag term. All the terms with the suffix `j' represents
jet quantities. If ${\cal{F}}_{r_j}={\cal{E}}_{r_j}={\cal{P}}_{r_j}=0$, then Eq. (\ref{radmjet.eq}) can be integrated to
give the Bernoulli equation for jet,
\begin{equation}
 {\cal E}_j=\frac{v^2_j}{2}+\frac{a^2_j}{\gamma -1}+\frac{\lambda_{j}^{2}}{2x_{j}^{2}}-\frac{1}{2(r_{j}-1)}
\label{jetbern.eq}
\end{equation}
For purely thermally driven flow ${\cal E}_j$ is a constant of motion, however, for a radiatively
driven flow ${\cal E}_j$ is a variable, as we will show later (Figs. 4e, 5e).
The jet outflow rate is given by
\begin{equation}
\dot{M}_{out}=\rho_{j}v_{j} \cal{A}, 
\label{jmf.eq}
\end{equation}
where ${\cal{A}}$ is the cross-sectional area of the jet.
Since absorption and emission terms in the jet is zero, therefore, it is described by
the polytropic equation of state
$(p_{j}=K_{j}\rho_{j}^{\gamma})$. 
The entropy accretion rate (from Eq. \ref{jmf.eq}) for the jet is given by
\begin{equation}
 {\dot {\cal M}}_j=a^{2n}_jv_j{\cal A}.
\label{entroj.eq}
\end{equation}
${\dot {\cal M}}_j$ is a measure of entropy of the jet, and is constant along the jet since
heating and cooling is ignored.
Equations~(\ref{radmjet.eq},\ref{entroj.eq} ) are simplified,
to obtain,
\begin{equation} 
\frac{dv_{j}}{dr}=\frac{\cal N}{\cal D},
\label{gvjet.eq}
\end{equation}
where,
\begin{equation}
{\cal N}=\frac{1}{2(r_j-1)^{2}}\frac{dr_j}{dr}
-\frac{\lambda_{j}^{2}}{x_{j}^{3}}\frac{dx_{j}}{dr}
-\frac{a_{j}^{2}}{{\cal{A}}}\frac{d{\cal{A}}}{dr}
-{\cal{F}}_{r_j}+v_j({\cal{E}}_{r_j}+{\cal{P}}_{r_j}),
\end{equation}
and,
\begin{equation}
 {\cal D}=\frac{a_{j}^{2}}{v_{j}}-v_{j}.
\end{equation}

The sonic point ($r_{jc}$) condition for jet is obtained for the condition 
$r_j\rightarrow r_{jc}$, $dv_j/dr\rightarrow 0/0$,
\begin{equation}
v_{jc}=a_{jc}=\frac{-A_2+\sqrt{A_{2}^2-4A_1A_3}}{2A_1};
\label{jetc.eq}
\end{equation}
where,
$$
A_1=\left[\frac{1}{{\cal{A}}_{c}}\left(\frac{d{\cal{A}}}{dr}\right)_{r_{c}}  \right],~~
A_2=-({\cal{E}}_{rj}+{\cal{P}}_{rj}),  ~~
A_3=\left[{\cal{F}}_{rj}-\frac{1}{2(r_{jc}-1)^{2}}\left(\frac{dr_{j}}{dr}\right)_{r_{c}}
+\frac{\lambda_{jc}^{2}}{x_{jc}^{3}}\left(\frac{dx_{j}}{dr}\right)_{r_{c}}  \right].
$$
The gradient of jet velocity on the sonic point is obtained by L'Hospital rule
\begin{equation}
\left[ \frac{dv_j}{dr}\right]_c=\left[\frac{d{\cal N}/dr}{d{\cal D}/dr}\right]_c.
\label{dvjc.eq}
\end{equation}

To obtain the jet solutions one has to integrate Eq. \ref{gvjet.eq}, with the help of Eqs. (\ref{jetc.eq}-\ref{dvjc.eq}).
In this paper we have chosen the adiabatic index to be $\gamma=1.4$, and has been shown to be the typical value
close to the black hole \citep{cc11}.

\subsection{The method to calculate self-consistent accretion-ejection solution}

The jet solution is dictated by the accretion solution, and now we discuss how we obtain the self-consistent
accretion-ejection solution. The accretion solution depends on the following constants of motion
$E$, $\lambda_0$,
and the viscosity parameter $\alpha$. Therefore, our first step is to obtain the accretion solution. 

Step 1: Initially we assume no jet \ie ${\dot M}_{out}=0$ and integrate Eqs. (\ref{rme.eq}-\ref{ege.eq}) \ie
Eqs. (\ref{grad.eq})
outwards and find the inner sonic point ($x_{ci}$) iteratively by choosing appropriate
$\dot{\cal{M}}$ and simultaneously checking for the sonic point conditions (Eqs. 16-19, of Kumar \& Chattopadhyay 2013).
The first hurdle to obtain the accretion solution is that although flow parameters are regular but there is a
coordinate singularity
on the horizon. However, the asymptotic values of $u(x)$, and $\lambda(x)$ very close to the horizon at $x_{in}$, are
obtained for appropriate values of ${\dot {\cal M}}$ \citep{bdl08}, and are given by,

\begin{equation}
\lambda(x)=\lambda_{0}\left[1+\frac{2\alpha}{\gamma r_{g}}\left(\frac{2}{r_{g}}\right)^{1/2}
\left(\frac{{\dot{\cal{M}}}^{2}}{2r^3_g}  \right)^{\frac{\gamma-1}{\gamma+1}}(x-r_g)^{\frac{\gamma+5}{2\gamma+2}}\right],
\ x\rightarrow 1 %r_{g}
\label{asam.eq}
\end{equation}
and
\begin{equation}
u(x)=u_{ff}(x)\left[1+\frac{2Ex^{2}-\lambda_{0}^{2}-(\gamma+1)f(x)}{x^{2}u_{ff}^{2}(x)-(\gamma-1)f(x)}  \right]^{1/2},
\ x\rightarrow 1 %r_{g}
\label{au.eq}
\end{equation}
where the function $f(x)$ is
$f(x)=2x^{2}(\gamma^{2}-1)^{-1}\left[{\dot{\cal{M}}}^{2}/\{2x^{3}(x-1)\} \right]^{\frac{\gamma-1}{\gamma+1}}$
and the free fall velocity in the pseudo-Newtonian potential is given by
$u_{ff}(x)=1/{\sqrt{(x-1)}}$. 
Without any loss of generality we choose $x_{in}=1.01$, and with these asymptotic values at $x_{in}$ we start the integration
of Eqs. \ref{grad.eq} and find the sonic points of the accretion solution. All kinds of solutions can be obtained,
ranging from solutions passing through one sonic point (ADAF type and for low angular momentum flow BONDI type, see
Kumar \& Chattopadhyay 2013), to flow solutions harbouring multiple sonic points. Only flows harbouring multiple sonic points
can have shocks. Once a flow passes through inner type sonic points $x_{ci}$, we integrate the flow equations
while checking for 
the shock conditions (Eq. \ref{rmf.eq}-\ref{ef1.eq}) and sonic point conditions to calculate 
shock location ($x_s$) and outer sonic point ($x_{co}$) simultaneously.
The shock conditions are,
\begin{equation}
W_{+}+\Sigma_{+} u_{+}^{2}=W_{-}+\Sigma_{-} u_{-}^{2},
\label{rmf.eq}
\end{equation}
%and angular momentum flux
\begin{equation}
E_{+} =E_{-},
\label{ef.eq}
\end{equation}
and the mass flux
\begin{equation}
{\dot M}_{+} ={\dot M}_{-}-{\dot M}_{out}={\dot M}_{-}\left(1-R_{\dot m}\right),~~ \mbox{where }
R_{\dot m}=\frac{{\dot M}_{out}}{{\dot M}_{-}}=\frac{R \rho_j v_j(x_b) {\mathcal A}(x_s)}{2 \pi \Sigma_+ u_-x_s},
\label{ef1.eq}
\end{equation}
where $+$ and $-$ represents the post-shock and pre-shock quantities, respectively and $W$
is vertically integrated pressure. Here, $R=u_-/u_+=\Sigma_+/\Sigma_-$ is the compression
ratio at the shock. If there is no shock then $R=1$, and otherwise $R>1$. As has been mentioned above,
the initial accretion solution is obtained by putting $R_{\dot m}=0$.  

Step 2: After we obtain the accretion temperature and number density distribution
from the accretion solution,
we compute the radiative moments  ${\cal E}_{rj}$, ${\cal{F}}_{r_j}$, and ${\cal P}_{r_j}$
as is discussed in section 2.2.1. These variables are used in jet equation.

Step 3: The extra compression in the post-shock flow drives bipolar outflows,
and the entire post-shock disc participates in jet generation \citep{mlc94,mrc96,ny09,nrks05}.
The jet base is the post-shock disc, so the jet is launched with the post-shock disc quantities.
If $x=x_b$ is the jet launch site in the post-shock disc, then the 
specific energy of the disc at $x_b$ \ie ${\cal E}_b={\cal E}(x_b)$, and the post-shock
specific angular momentum at $x_b$ \ie $\lambda_b=\lambda (x_b)$ and the density
at the top of the disc at $x_b$ \ie  $\rho_b$, are the flow variables of the jet at its base. 
Here we have taken $x_b=(x_{ci}+x_s)/3$, and since the disc
is assumed to be in 
hydrostatic equilibrium along the $z$ direction, $\rho_b=\rho(x_b){\rm exp}(-2y_b/h_s)$,
where, $y_b$ is the height of the disc at $x_b$, and $h_s$ is disc height at shock location.
The jet velocity $v_{jb}$ and the sound speed $a_{jb}$ at the jet base are not arbitrarily
assigned, but are computed self consistently. We eliminate $a_j$ from Eq. (\ref{entroj.eq}) and Eq. (\ref{jetbern.eq}),
and express ${\cal E}_{b}$ in terms of ${\dot {\cal M}}_{j}$, $v_{jb}$, $\lambda_{b}$ etc. Since ${\cal E}_{b}$
and $\lambda_{b}$ are obtained from the post-shock disc and therefore known for the jet, we iterate
with various values of entropy \ie ${\dot {\cal M}}_{j}$ to obtain the correct value of $v_{jb}$ and $a_{jb}$, with which 
the unique transonic solution is determined by
integrating Eq. (\ref{gvjet.eq}) and checking for the sonic point conditions Eqs. (\ref{jetc.eq},\ref{dvjc.eq}).
Once the transonic jet solution is obtained, then $R_{\dot m}$ is easily calculated from Eq. (\ref{ef1.eq}).

Step 4: Now the
information of $R_{\dot m}$ is used in the shock condition (Eqs. \ref{rmf.eq} - \ref{ef1.eq})  
and then go back to Step 1 to recalculate the new accretion solution, the new radiative
moments (Step 2), and the new jet solution (Step 3). This process is repeated till the
shock location converges to a value ($x_s$), and we
find out the self-consistent, converged accretion-ejection solution.  

Therefore, there are three iteration processes, namely, (1) to find out inner sonic point of accretion disc, (2)
accretion shock and the outer sonic point, and (3) the jet sonic point when launched with the 
post-shock disc variables.

\subsubsection{Radiative moments}

\begin{figure}
 \begin{center}
\epsfig{figure=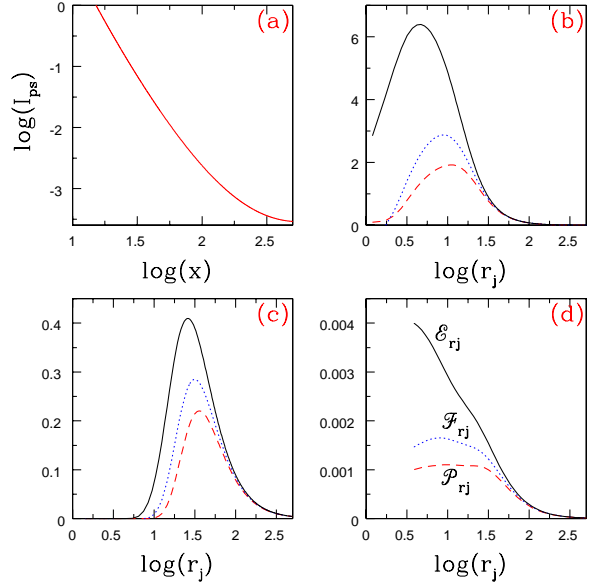,height=8.cm,width=8.cm,angle=0}
\caption{(a) Normalized intensity of synchrotron radiations (solid line, online red) from pre-shock disc
with $log(x)$ (b) Moments of the radiation field due to the post-shock disc
${\cal E}_{r_js}/{\cal K}_0$ (solid, online black), ${\cal F}_{r_js}/{\cal K}_0$ (dotted,
online blue) and
${\cal P}_{r_js}/{\cal K}_0$ (dashed, online red),
are plotted with $log(r_j)$, and
(c) ${\cal E}_{r_jps}/{\cal J}_0$ (solid, online black), ${\cal F}_{r_jps}/{\cal J}_0$ (dotted,
online blue) and
${\cal P}_{r_jps}/{\cal J}_0$ (dashed, online red), the moments of the radiation field due to the pre-shock disc are plotted.
(d) Total moments of radiation field due to post-shock and pre-shock disc,
are generated by an accretion disc with
parameters $E = 3.5\times10^{-3}$, $\lambda_0 = 1.5435$, $\alpha = 0.02$, $\gamma = 1.4$ (forming
shock at $x_s=15.18$). The radiation field is for equal post-shock and pre-shock luminosities
${\ell}_s = {\ell}_{ps} = 0.4$.}
 \end{center}
\label{lab:fig2}
\end{figure}

Now as the jet is generated due to compression as well as shock heating from the post-shock disc
(see Fig. \ref{lab:fig1}), the jet will be moving through the radiation field
of the accretion disc. We now outline the method to compute radiative moments and the net radiative acceleration
of the jet.
The radiative acceleration term is proportional to the radiative flux $F_{r_j}$, 
but the radiation drag term (negative terms in r. h. s of Eq. \ref{radmjet.eq}) depends on jet 
flow velocity $v_j$, the radiation energy density $E_{r_j}$ and the 
radiation pressure $P_{r_j}$. Figure \ref{lab:fig1} shows the schematic diagram
of the radiations coming from various parts of the disc, and how it can interact
with the outflowing jet, generated due to shock heating of the inner disc. 
The radiative moments along the jet streamlines ($r_j$) are calculated from the post-shock ($s$)
and pre-shock ($ps$) disc respectively. The radiative terms in Eq. \ref{radmjet.eq} are given by,
\begin{equation}
 {\cal E}_{r_j}=\frac{\sigma_T}{mc}\left(\int I_s d\Omega_s + \int I_{ps} d \Omega_{ps}\right)=
\frac{\sigma_T}{m}\left(E_{r_js}+E_{r_jps} \right)={\cal E}_{r_js}+{\cal E}_{r_jps}
\end{equation}
\begin{equation}
 {\cal F}_{r_j}=\frac{\sigma_T}{mc}\left(\int I_s {\hat {r_j}} d\Omega_s + \int I_{ps}{\hat {r_j}}
d \Omega_{ps}\right)=\frac{\sigma_T}{mc}\left(F_{r_js}+F_{r_jps}\right)={\cal F}_{r_js}+{\cal F}_{r_jps}
\end{equation}
\begin{equation}
 {\cal P}_{r_j}=\frac{\sigma_T}{mc}\left(\int I_s {\hat {r_j}}{\hat {r_j}} d\Omega_s + \int I_{ps}{\hat {r_j}}
{\hat {r_j}} d \Omega_{ps}\right)=\frac{\sigma_T}{m}\left(P_{r_js}+P_{r_jps}\right)={\cal P}_{r_js}+{\cal P}_{r_jps}
\end{equation}
Since the jet streamline is close to the axis of symmetry, we calculate the radiative moments on the axis and approximate
these to hold at the same radial distance on the jet streamline.
In this paper, we are not dealing with the detailed features of the
radiation spectrum from the accretion disc. Rather, we are interested to see the effect 
of the total pre-shock and post-shock radiation on the acceleration of the jet.
Hence, we do not include radiative transfer dynamically into
the hydrodynamic solution. To calculate the moments of radiation due to the pre-shock disc, 
as an example, we consider
only synchrotron processes from the pre-shock disc.

The synchrotron emissivity is due to the presence of  stochastic magnetic
field, where the magnetic pressure ($p_m$) is in partial equipartition with the gas pressure ($p_g$), \ie
$$
p_m=\frac{B^2}{8\pi}=\beta p_g; ~~\mbox{where,}~  0 \leq \beta \leq 1,
$$
here, for $\beta=0$ implies no magnetic field and therefore the total pressure in Eq. (\ref{rme.eq}) $p=p_g$,
while $\beta=1$ implies strict equipartition between gas and magnetic pressure and therefore
$p=p_g+p_m$.
The analytical expression for synchrotron emissivity is given by \citep{st83}, and the resulting intensity
is,
\begin{equation}
I_{ps}=I_{syn}=\frac{16}{3}\frac{e^2}{c}\left( \frac{eB}{m_e c} \right)^2 \Theta^2 n_e \left(\frac{h r_g}{sec{\theta}_{ps}}
\right) \ \  {\rm erg} \
{\rm cm}^{-2} {\rm s}^{-1}
\label{syn.eq}
\end{equation}
where, $\Theta, n_e, h$ and $\theta_{ps}$ are the pre-shock local dimensionless temperature $k_b T/(m_e c^2)$,
electron number
density, disc half height and angle from the axis of symmetry to the pre-shock disc surface, respectively. The dependence of
$I_{ps}$ on disc radius is through flow variable like $T~ \&~ n_e$. Integrating $I_{ps}$ over 
the pre-shock disc gives us the pre-shock luminosity $\ell_{ps}$. 
The radiative intensity from the post shock disc is $I_s={\ell_s}/{\sc A}_s$, where
$\ell_s$ is the post-shock luminosity in units of Eddington luminosity and
${\sc A}_s$ is the total surface area of the post-shock disc.
The total luminosity is $\ell = \ell_s+ \ell_{ps}$.
The moments of the radiation field above the accretion disc was calculated before 
\citep{cc00,cc02,cdc04,c05}, and for the above mentioned approximations, they are given by,
\begin{equation}
{\cal E}_{r_js}=2\pi{\cal K}_0\int^{x_s}_{x_{in}}\frac{zxdx}
{[(z-x~cot\theta_s)^2+x^2]^{3/2}}; ~~
{\cal E}_{r_jps}=2\pi{\cal J}_0\int^{x_{inj}}_{x_s}\frac{a^5zxdx}
{u^2x^{3/2}(x-1)[(z-x~cot\theta_{ps})^2+x^2]^{3/2}} ,
\end{equation}
\begin{equation}
{\cal F}_{r_js}=2\pi{\cal K}_0\int^{x_s}_{x_{in}}\frac{z(z-x~cot\theta_s)xdx}{[(z-x~cot\theta_s)^2+x^2]^2}; ~~
{\cal F}_{r_jps}=2\pi{\cal J}_0\int^{x_{inj}}_{x_s}\frac{a^5z(z-x~cot\theta_{ps})xdx}
{u^2x^{3/2}(x-1)[(z-x~cot\theta_{ps})^2+x^2]^2},
\end{equation}
\begin{equation}
 {\cal P}_{r_js}=2\pi{\cal K}_0\int^{x_s}_{x_{in}}\frac{z(z-x~cot\theta_s)^2xdx}{[(z-x~cot\theta_s)^2+x^2]^{5/2}}; ~~
{\cal P}_{r_jps}=2\pi{\cal J}_0\int^{x_{inj}}_{x_s}\frac{a^5z(z-x~cot\theta_{ps})^2xdx}
{u^2x^{3/2}(x-1)[(z-x~cot\theta_{ps})^2+x^2]^{5/2}},
\end{equation}

\begin{figure}
\begin{center}
 \epsfig{figure=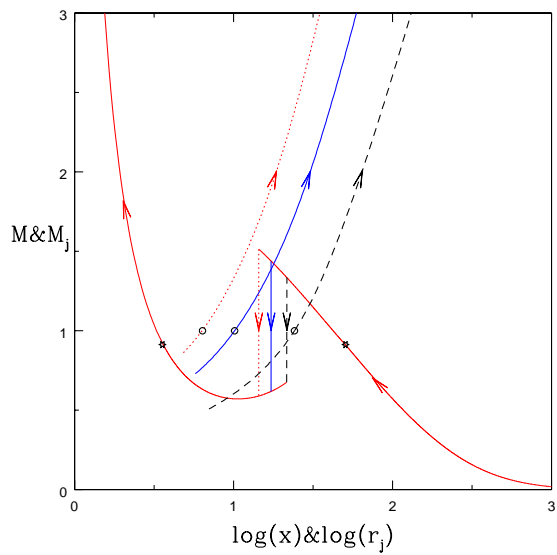,height=8.cm,width=8.cm,angle=0} 
\caption{Accretion Mach number ($M$) and jet Mach number ($M_j$)
with radial distance $x$ and jet streamline $r_j$
in log scale, are shown in the plot. Open circle represents the critical points of
jets and star marks for accretion solutions, while arrows show flow directions.
Solid curve (red online) with inward arrows
is accretion solution for parameters $E = 3.3\times10^{-3}, ~ \lambda_0
=1.353,~ \alpha = 0.1$. The post
shock disc generate jets.
Each jet curve represents $M_j$ vs $log(r_j)$ for parameters $\ell_s =  0.0~(\ell =0.0386)$ (dashed, online black),
$\ell_s = 0.1~(\ell = 0.151)$ (solid, online blue), and  $\ell_s = 0.2~(\ell = 0.274)$ (dotted,
online red). The shocks are shown by vertical jumps (line style and color corresponds to that of the jet it generates).
The parameter $\beta {\dot m}^2= 0.01$ is kept constant.}
\end{center}
\label{lab:fig3}
\end{figure}

\noindent where, $\theta_s=tan^{-1}(x_s/h_s)$, $h_s=\sqrt{(2/\gamma)}a_{s+}x^{1/2}_s(x_s-1)$ and
$\theta_{ps}=tan^{-1}(x/h)$ for pre-shock disc, \ie for $x>x_s$.
Moreover, 
\begin{eqnarray}
&& {\cal K}_0=\frac{1.3{\times}10^{38}{\ell}_s{\sigma}_{T}}{2{\pi}cm_p{\sc A}_sGM_{\odot}} 
\label{radcons1.eq} \\ 
&& \mbox{ and, }
{\cal J}_0=\frac{2.93\times 10^{34}e^4\mu^2\beta\sigma_T{\dot m}^2}{3\pi m^4_ec^2 sec\theta_{ps}\gamma^{5/2}G^2M^2_{\odot}}
\label{radcons2.eq}
\end{eqnarray}
where, ${\dot m}$ is the accretion rate in units of Eddington accretion rate,
$\sigma_T$ is Thomson scattering cross section, $\mu$ is mean molecular weight of the plasma,
$m_p$ is the proton mass, $m_e$ is the electron mass and $M_{\odot}$ is the solar mass. It is to be noted that
the pre-shock radiation would depend on the product $\beta {\dot m}^2$. 
It is interesting to know that, the post-shock region by virtue of its geometry will block some of the pre-shock
photons to the base of the jet, an effect coined as the shadow effect of the post-shock disc \citep{cdc04,c05}.
That is to say, if the height of the jet is $y_j<y_{jl}$, then ${\cal E}_{r_jps}={\cal F}_{r_jps}={\cal P}_{r_jps}=0$,
where 
\begin{equation}
 y_{jl}=h_s-\frac{h_{inj}-h_s}{x_{inj}-x_s}\left(x_s-x_j\right),
\end{equation}
here, $h_{inj}$ \& $x_{inj}$ are height and radius at outer edge
of the disc. Additionally due to the shadow effect, the inner edge of the pre-shock
disc as seen by an observer at some height $y_j$ will be,
$$
x_i=-\frac{x_s y_j}{h_s-y_j-x_s cot\theta_{ps}}
$$
In this paper, we use $\beta{\dot m}^2$ as a supplied parameter to calculate $\ell_{ps}$. 
Also we use $\ell_s$ initially as a parameter (except Fig. 8 and  Fig. 9) in order to understand the 
effect of the relative proportions of radiation
coming from different parts of disc on the ejected jet, but finally we will use the relation
${\ell}_s/{\ell}_{ps}$ as a function
of $x_s$ (see Appendix), to compute ${\ell}_s$
from the pre-shock radiation, and solve the accretion-ejection solution.

In Figs. 2a-d, all the plots are generated for a disc
characterized by $(E,\lambda_0,\alpha, \gamma )=(3.5\times10^{-3},1.5435,0.02, 1.4)$, 
which generates a shock at $x_s=15.18$.
In Fig. 2a, $I_{ps}$ is plotted as a function of $x$, starting from $x_s$ outwards.
In Fig. 2b, we plot ${\cal E}_{r_js}/{\cal K}_0$ (solid, online black),
${\cal F}_{r_js}/{\cal K}_0$ (dotted, online blue) and
${\cal P}_{r_js}/{\cal K}_0$ (dashed, online red), the radiation moments due to the post-shock disc.
In Fig. 2c,
we plot ${\cal E}_{r_jps}/{\cal J}_0$ (solid, online black), ${\cal F}_{r_jps}/{\cal J}_0$
(dotted, online blue) and
${\cal P}_{r_jps}/{\cal J}_0$ (dashed, online red), the radiation moments due to the pre-shock disc. 
The shadow effect
of the post-shock disc is clearly shown in the figure. In Fig. 2d,
we plot ${\cal E}_{r_j}$ (solid, online black), ${\cal F}_{r_j}$
(dotted, online blue) and ${\cal P}_{r_j}$ (dashed, online red)
taking equal post-shock and pre-shock luminosities, ${\ell}_s = \ell_{ps} = 0.4$, respectively.
From the above figures it is clear that radiative moments from the post shock region peaks at a 
height closer to the black hole
and the moments from the pre-shock disc peaks typically at a distance few times larger. 
However, the space dependent parts
of the moments (Figs. 2b-c) shows that the moments of radiation from the pre-shock disc is typically
an
order of magnitude weaker than those due to the post-shock disc. However, the second distinct bulge
 in Fig. 2d shows, if the pre-shock luminosity
is comparable to post-shock luminosity then the radiation moments peaks at two places, and hence
presents a prospect for multi-stage acceleration scheme for the jets. 
%%%%%%%%%%%%%%%%%%%SECTION 3 %%%%%%%%%%%%%%%%%%
\section{Results}

In \citet{kc13} we have shown in details various
cases of accretion solutions, and also have shown that post shock disc naturally produces bipolar outflows.
In this paper, therefore we discuss only the shocked accretion solution.
The accretion solutions are characteristic by flow parameters like the grand energy $E$, the specific angular
momentum at the horizon $\lambda_0$ (conversely, $\lambda_{inj}$ at the outer edge of the disc) and the viscosity parameter
$\alpha$. The jet is launched from the disc with the specific energy (${\cal E}_j$ and angular momentum ($\lambda_j$)
of the post-shock disc at the jet launch site ($r_b$). We also compute the moments of radiation fields (${\cal E}_{r_j}$,
${\cal F}_{r_j}$, ${\cal P}_{r_j}$) above the disc,
and are used to accelerate the jets. The pre-shock disc radiation depends on $\beta{\dot m}^2$, where, $\beta$ is the ratio of the
magnetic and the gas pressure and ${\dot m}$ is the accretion rate in units of Eddington rate. In order to find the
effect of post-shock radiation and pre-shock radiation on jet acceleration, the post-shock luminosity $\ell_s$ and
$\beta{\dot m}^2$ (conversely, $\ell_{ps}$ the pre-shock luminosity) are supplied as independent parameters. However, since
post-shock radiation is produced by inverse-Comptonization of self generated and intercepted photons from the pre-shock
disc, we compute the post-shock radiation self-consistently by employing the techniques of \citet{cm06}.
And in Figs. (8, 9, 10) the jets are accelerated by self consistent estimation of radiation field,
both from pre and post-shock discs.
\subsection{Effect of post-shock radiation on jet acceleration}
The radiations produced from accretion is governed by the solution. However, we will now treat the post-shock luminosity
as a parameter just to see how it affects the jet acceleration.
The accretion Mach number $M=u/a$ with $log(x)$ is plotted
for the parameters $E = 3.3\times10^{-3},~ \lambda_0
=1.353,~ \alpha = 0.1$, in Fig. 3. We first consider $\ell_s=0$ and $\beta{\dot m}^2=0.01$,
the shock is at $x_s=21.546$ (vertical dashed, online black), and drives bipolar outflow, where
the jet Mach number $M_j$ (dashed, online black)
is plotted with $log(r_j)$, and the mass outflow rate is
$R_{\dot{m}}=0.032$ and the sonic point of the jet is at $r_{jc}=24.160$.
Radiation field from the accretion disc deposit its momentum and accelerate the shock generated jets.
%Assuming a radiation field for , and 
Assuming $\ell_s=0.1$ for same set of accretion boundary condition,
accretion-ejection solution is computed.
The shock is found to be at $x_s=17.92$ (solid, online blue) 
and the jet $M_j$ distribution (solid, online blue) has a sonic point at $r_{jc}=9.59$ and $R_{\dot m}=0.046$.
The total luminosity for this case is $\ell = (\ell_s+\ell_{ps})=0.151~(\ell_s = 0.1)$.
Keeping $\beta{\dot m}^2$ same, we increase $\ell_s=0.2$, so the total luminosity increases to $\ell = 0.274$,
The accretion-ejection solution shows that the mass outflow increases to $R_{\dot m}=0.058$
and the shock in accretion decreases to $x_s=14.4$ (dotted, online red). In this case too,
due to the increase of radiation, the sonic point decreases to $r_{jc}=6.36$,  indicating stronger jet.
It is interesting to note that, the pre-shock luminosity
increases even though $\beta{\dot m}^2$ is kept constant, because, with the decrease in $x_s$,
the size of the pre-shock disc increases.

\begin{figure}
\begin{center}
 \epsfig{figure=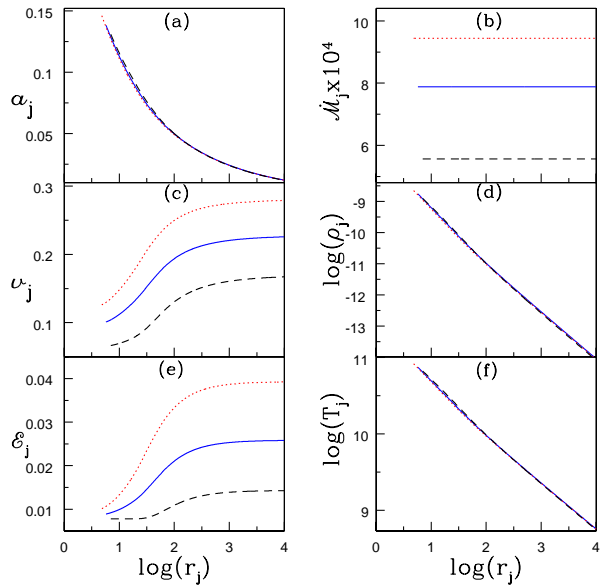,height=8.cm,width=8.cm,angle=0} 
\caption{Jet flow variables like (a) sound speed $a_j$, (b) entropy-accretion rate $\dot{\cal{M}}_j$,
(c) jet velocity $v_j$, (d) density $\rho_j$,
(e) specific energy ${\cal{E}}_j$, and (f) temperature $T_j$ are plotted with $log(r_j)$.
The disc, radiation parameters, the line colours and line styles are same as Fig. 3.}
\end{center}
\label{lab:fig4}
\end{figure}

The jet solutions are explored in more details in the following figure.
Jet variables $a_j$ (Fig. 4a), $\dot{\cal{M}}_j$ (Fig. 4b), $v_j$ (Fig. 4c), $\rho_j$ (Fig. 4d), ${\cal E}_j$
(Fig. 4e) and $T_j$ (Fig. 4f) are plotted
with $log(r_j)$. Each curve corresponds to $\ell_s=0$ (dashed, online black), $\ell_s=0.1$ (solid, online blue),
and $\ell_s=0.2$ (dotted, online red),
and which are exactly the same cases of jet solutions as in Fig. 3.
The increase in ${\ell}_s$, accelerates the jets further, and therefore increases $R_{\dot m}$, this in turn
decreases the post-shock pressure and the location of the shock front moves close to the horizon. As a result
the jet base
moves closer. 
At a given $r_j$, we find $v_j$ is higher for higher $\ell_s$, but difference in $a_j$ or $T_j$ are imperceptible.
This shows, that enhanced jet acceleration is due to the radiative momentum deposition onto the jets
and not due to conversion of thermal energy to the kinetic one. Since higher $\ell_s$ accelerate the jet,
the sonic point is formed closer to the horizon. Higher $\ell_s$ not only means faster jet, but also a jet with higher
entropy (Fig. 4b).

\begin{figure}
 \begin{center}
\epsfig{figure=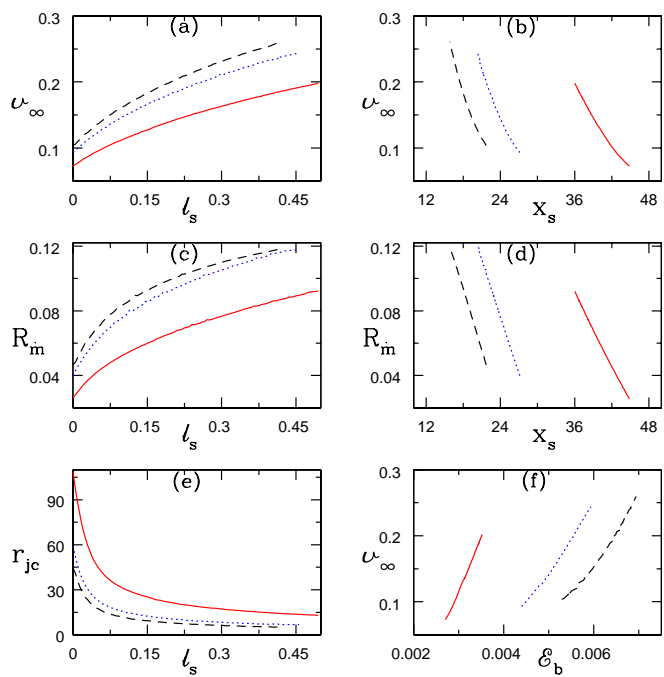,height=9cm,width=9cm,angle=0}
\caption{Jet terminal velocity $v_{\infty}$ is plotted with $\ell_s$ (a), with $x_s$ (b), and with ${\cal E}_b$ (f).
$R_{\dot m}$ is plotted with $\ell_s$ (c) and $x_s$ (d). Moreover, $r_{jc}$ is plotted with $\ell_s$
(e) too.
The accretion solution corresponds to disc parameters $E=0.001$ and $\lambda_{inj}=70.6$ at disc outer boundary
$x_{inj}=10^4$
and keeping pre-shock luminosity is $\ell_{ps}=0.0$. Each curve correspond for $\alpha=0.04944$ (solid, online
red),
$0.049887$ (dotted, online blue),
and $0.050113$ (dashed, online black).}
 \end{center}
\label{lab:fig5}
\end{figure}

The jet terminal speed is defined as $v_{\infty}=v~({\rm{at}}~r_j\rightarrow {\rm large})$ where $dv_j/dr_j\rightarrow 0$.
As shown in Figs. 4, $v_{\infty}$ increases
appreciably with the increase of $\ell_s$ and $\ell_{ps}$, for a given value of disc viscosity parameter
${\alpha}$. We would like to see whether this behaviour of $v_{\infty}$ holds true for a range of
$\alpha$. In Fig. 5a, $v_{\infty}$ is plotted with $\ell_s$, in Figs. 5b \& 5f, $v_{\infty}$ is plotted with $x_s$
and the jet base specific energy ${\cal E}_b$. And $R_{\dot m}$ is plotted with $\ell_s$ (Fig. 5c) and $x_s$
(Fig. 5d). Each curve correspond for $\alpha=0.04944$ (solid, online
red),
$0.049887$ (dotted, online blue),
and $0.050113$ (dashed, online black).
All these figures are generated with outer
boundary condition $E=0.001$ and $\lambda_{inj}=70.6$ at $x_{inj}=10^4$.
It is clear that $v_{\infty}$ increases with ${\ell}_s$ at a given value of $\alpha$, as well as,
increases with $\alpha$ at a given $\ell_s$. Since $x_s$ moves closer with the increase of both $\ell_s$ and 
$\alpha$, the jets are launched with higher ${\cal E}_b$, which in turn increases $v_{jb}$. This is also the
reason that the relative mass outflow rate $R_{\dot m}$ increases with the increase of both $\ell_s$ and $\alpha$.
Since $v_{jb}$ increases with $\ell_s$ and $\alpha$, therefore the jets become supersonic at a distance nearer to the
jet base,
\ie $r_{jc}$ decreases with the increase of $\ell_s$ (Fig. 5e).

\subsection{Effect of pre-shock radiation on jets}
\begin{figure}
\begin{center}
 \epsfig{figure=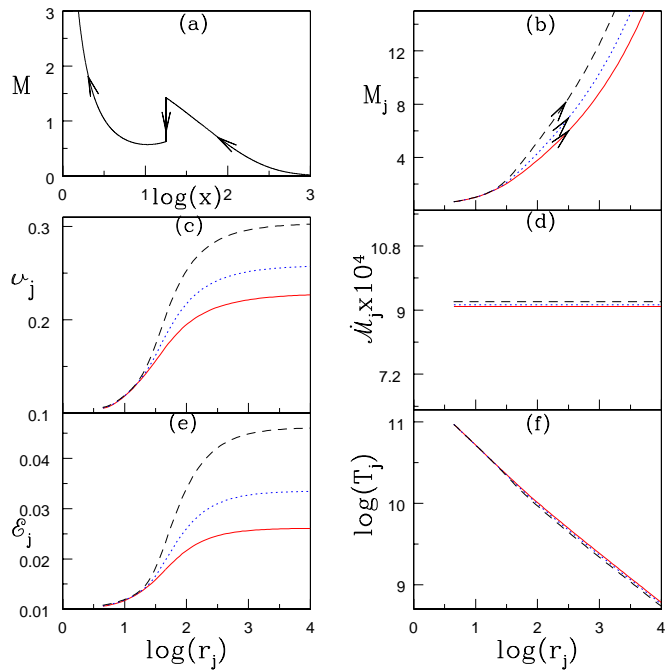,height=9.cm,width=9.cm,angle=0} 
\caption{(a) Accretion Mach Number $M$ with $log(x)$, and
 (b) jet Mach number $M_j$, (c) jet velocity $v_j$, (d) jet entropy rate ${\dot{\cal{M}}}_j$,
(e) specific energy ${\cal{E}}_j$, and (f) temperature $T_j$ plotted with $log(r_j)$
for the disc parameters $E=3.3\times10^{-3}, \lambda_0=1.353,
\alpha=0.1$,
and shock forms at $x_s=17.92$. Each curve is for $\beta{\dot m}^2=0.01~(\ell=0.151)$ (solid, online red),
$\beta{\dot m}^2=0.02~(\ell=0.202)$ (dotted, online blue), and $\beta{\dot m}^2=0.04~(\ell=0.303)$
(dashed, online black) and $\ell_s=0.1$ 
is kept constant. 
Each jet solutions have critical points at $r_{jc}\sim 9.59$ and mass outflow rates, $R_{\dot{m}}\sim 0.046$.}
\end{center}
\label{fig6}
\end{figure}

The effect of radiation from pre-shock disc as it impinges on the jet is illustrated through Figs. 6b-f. In Fig. 6a,
the Mach number $M$ of the accretion solution is plotted with $log(x)$ for disc parameters $E=3.3\times10^{-3},
\lambda_0=1.353, \mbox{ and }\alpha=0.1$. The shock is at $x_s=17.92$. We choose $\ell_s=0.1$, but
the pre-shock radiation is changed by varying $\beta{\dot m}^2$. The solutions correspond to
$\beta{\dot m}^2=0.01~(\ell=0.151)$ (solid, online red),
$\beta{\dot m}^2=0.02~(\ell=0.202)$ (dotted, online blue), and $\beta{\dot m}^2=0.04~(\ell=0.303)$
(dashed, online black).
The post-shock disc
actually hides the base of the jet from most of the radiation from the pre-shock disc, while shines its own light
onto the jets (see Fig. \ref{lab:fig1}). As a result, if the post-shock radiation remains unaltered and the jet sonic point
is formed in the portion of the funnel like region where pre-shock radiation is negligible, then the jet base velocity
$v_{jb}$, the jet base $r_{jb}$ or the jet base properties are likely to remain roughly same, keeping the massloss
rate unaltered. Consequently,
the change in the accretion shock is imperceptible (Fig. 6a). All the jet variables closer to the base,
\eg $M_{j}$ (Fig. 6b), $v_{j}$ (Fig. 6c), ${\cal E}_{j}$ (Fig. 6e), and $T_{j}$ are indistinguishable, while
they differ from each other in the supersonic region, where the interaction of pre-shock radiation with the jet is
significant
too. It is to be remembered, that the temperature plotted here is the single temperature of the outflow.
The corresponding electron temperature should be about 2 orders of magnitude less.
However, the entropy accretion rates ${\dot {\cal M}}$ (Fig. 6d) are distinguishable even at the base.
Once again it is clear from the temperature plot, that radiative driving is significant.
It is also interesting to note from Figs. 3-4, that increasing $\ell_s$, would
result in faster jets, with higher $R_{\dot m}$ and lower jet sonic point ($r_{jc}$). 
While increasing the pre-shock radiation
also results in faster jets, but with almost no change in $R_{\dot m}$ and $r_{jc}$. Since relative mass outflow rates
affects the accretion solutions (Eq. \ref{ef1.eq}), so the feed-back effect of the jet on the disc due to
$\ell_s$ might be more significant than that due to ${\ell}_{ps}$.

\begin{figure}
 \begin{center}
\epsfig{figure=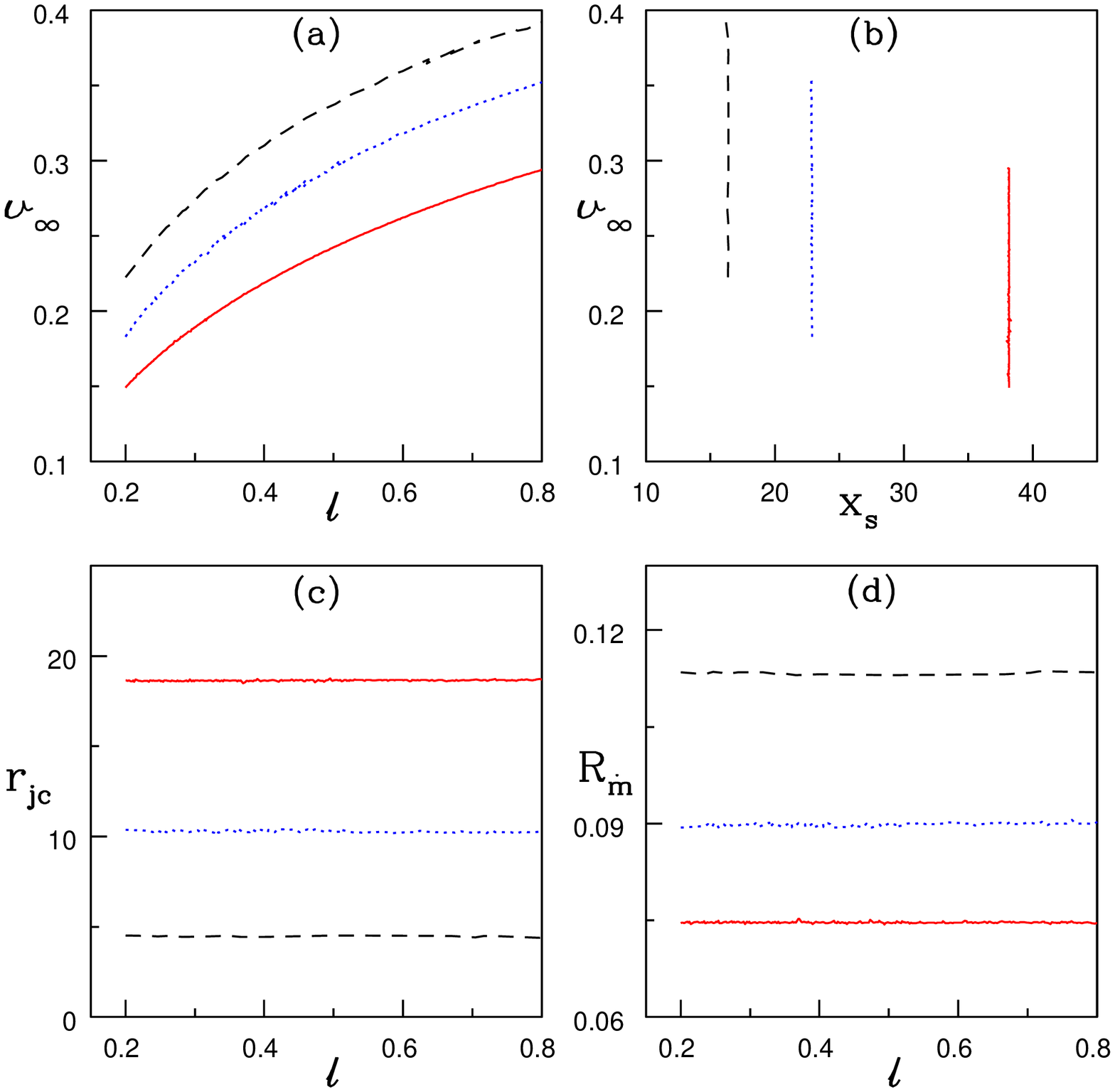,height=8.cm,width=8.cm,angle=0}
\caption{Terminal jet velocity $v_{\infty}$ with total disc 
luminosity $\ell$, (b) shock locations $x_s$, and (c) $r_{jc}$ is plotted with $\ell$
and (d) $R_{\dot m}$ is plotted with $\ell$. The 
disc parameters are $E=0.001$ and $\lambda_{inj}=70.6$ at disc outer boundary $x_{inj}=10^4$. and 
keeping post-shock luminosity is $\ell_{s}=0.2$ but varying pre-shock luminosity($\ell_{ps}$). 
Each curve are for viscosity parameter $\alpha=0.04944$ (solid, online red),
$0.049887$ (dotted, online blue), and $0.050113$
(dashed, online black).}
 \end{center}
\label{lab:fig7}
\end{figure}

In Figs. 7a-d, we investigate how $\ell_{ps}$
affects the jet solutions for a variety of $\alpha$, but for the same outer
boundary condition as in Fig. 6 with constant $\ell_s=0.2$.
In Fig. 7a, we plot $v_{\infty}$ as a function of $\ell$, where each curve represent disc
solutions with $\alpha=0.04944$ (solid, online red), $0.049887$ (dotted,
online blue), and $0.050113$ (dashed, online black).
In all these plots $\ell_{ps}$ varies from $0\rightarrow 0.6$.
Due to radiative driving, $v_{\infty}$ increases with $\ell$, and at a given $\ell$, it increases with
$\alpha$. As has been explained in connection to the previous figure, increasing $\alpha$ for a fixed
outer boundary decreases $x_s$, which means the jet base energy ${\cal E}_b$ increases, resulting in faster jet.
However, since the pre-shock disc primarily shines radiation on the supersonic part of the jet, therefore $\ell_{ps}$
has marginal effect on $r_{jc}$, and $v_{jb}$. Therefore, $R_{\dot m}$ is almost constant with the change of $\ell_{ps}$
(Fig. 7d),
which in turn keeps $x_s$ almost unchanged (Fig. 7b), and the jet sonic point $r_{jc}$ also remains
unchanged (Fig. 7c). In other words, we may conclude, that the radiation from the inner torus
of the accretion disc accelerate the jet, but also increases net mass-loss.
On the other hand,
radiation from pre-shock disc or the outer disc, accelerates the jet appreciably, although, has almost no effect on
$R_{\dot m}$.
Therefore, in the second case we may obtain jets with higher kinetic luminosity.
This conclusion is valid for any value of $\alpha$ which admits accretion shock.

\subsection{Radiative driving of jets with computed post-shock and pre-shock radiations}
\begin{figure}
 \begin{center}
  \epsfig{figure=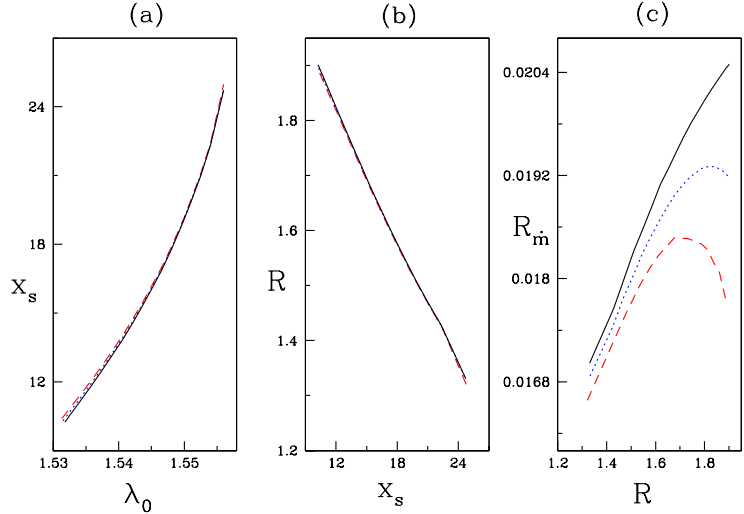,height=10.cm,width=10.cm,angle=0}
\caption{(a) Shock location $x_s$ is plotted with angular momentum at the horizon ($\lambda_0$),
(b)Compression ratio $R$ with $x_s$ and (c) mass outflow rate ($R_{\dot m}$) versus $R$
are plotted for flows with $E=0.004$, $\alpha=0.01$.
Various plots represent thermally driven flow (dashed, online red), and radiatively driven outflows 
for $\beta {\dot m}^2=0.005$ (dotted, online blue) and $\beta {\dot m}^2=0.01$ (solid,
online black).}
\end{center}
\label{lab:fig8}
\end{figure}

Having investigated the role the pre-shock and post-shock radiation may have on jets, we 
actually calculate the $\ell_s$ from $\ell_{ps}$. 
In appendix A, we have
discussed the different radiation processes in a general radiative transfer 
model (Chakrabarti \& Mandal 2006). From the accretion disc spectrum, 
we calculate the ratio of post-shock to 
pre-shock luminosity ($\ell_s/\ell_{ps}$) using the viscous transonic solution.
We then use a fitting formula of $\ell_s/\ell_{ps}$ (Eq. \ref{poly.eq}) to 
calculate the ratio at any given shock location ($x_s$). 
In (Fig. \ref{lab:figA2}1b), a typical ${\ell}_s/{\ell}_{ps}$ dependence 
on $x_s$ is obtained where the dots are the data points from model 
(Chakrabarti \& Mandal 2006) and solid line represents the fitting function.
We assume that the behaviour of this ratio with shock location is generic.
The procedure for calculating $\ell_{s}$ is as follows --- for a given set of values of $E,\lambda_0, \alpha$, the accretion
solution shows a shock at $x_s$ and a bipolar outflow with some $R_{\dot m}$.
We then calculate $\ell_{ps}$ (Eq. \ref{syn.eq}) by supplying $\beta {\dot m}^2$ and the density 
and temperature distribution between $x_{inj}$ and $x_s$. 
We use the fitting formula of $\ell_s/\ell_{ps}$
(Eq. \ref{poly.eq}) to calculate $\ell_s$. Using these the jet solution is obtained.
All the solutions presented in Figs. (3-7), we have solved the accretion ejection 
solution with the following fluid parameters $E$, $\lambda_0$ (at the horizon, or equivalently, $\lambda_{inj}$
at the outer boundary), $\alpha$, and in addition the radiation parameters $\ell_s$
and $\beta {\dot m}^2$ (equivalently $\ell_{ps}$).
Now, following the procedures described in appendix A, we reduce one parameter, namely, $\ell_s$.
In Fig. 8a, we plot $x_s$ with $\lambda_0$, in fig. 8b, we plot the compression ratio $R$ with $x_s$, and in fig. 8c,
mass outflow rate $R_{\dot m}$ with the compression ratio $R$. All the plots are for accretion disc
parameters $E=0.004$, $\alpha=0.01$, and various results has been obtained by varying $\lambda_0$.
The curves are for thermally driven jet (\ie $\beta \dot m^2=0$; dashed, online red), and 
thermal plus radiatively driven jets (\ie $\beta \dot m^2=0.005$; dotted online blue, and
$\beta \dot m^2=0.01$; solid, online black).
This shows that as the compression at the shock increases, it forces more matter into the
jet channel. Although $R$ increases as $x_s$ decreases, but smaller post-shock region means less matter can be
driven as jets, so $R_{\dot m}$ maximizes at some intermediate $R$. It has also been shown earlier \citep{c99,dcnc01}, that
for $R=1$ \ie no shock, $R_{\dot m}\sim 0$ \ie for no shock there is no outflow.

\begin{figure}
 \begin{center}
  \epsfig{figure=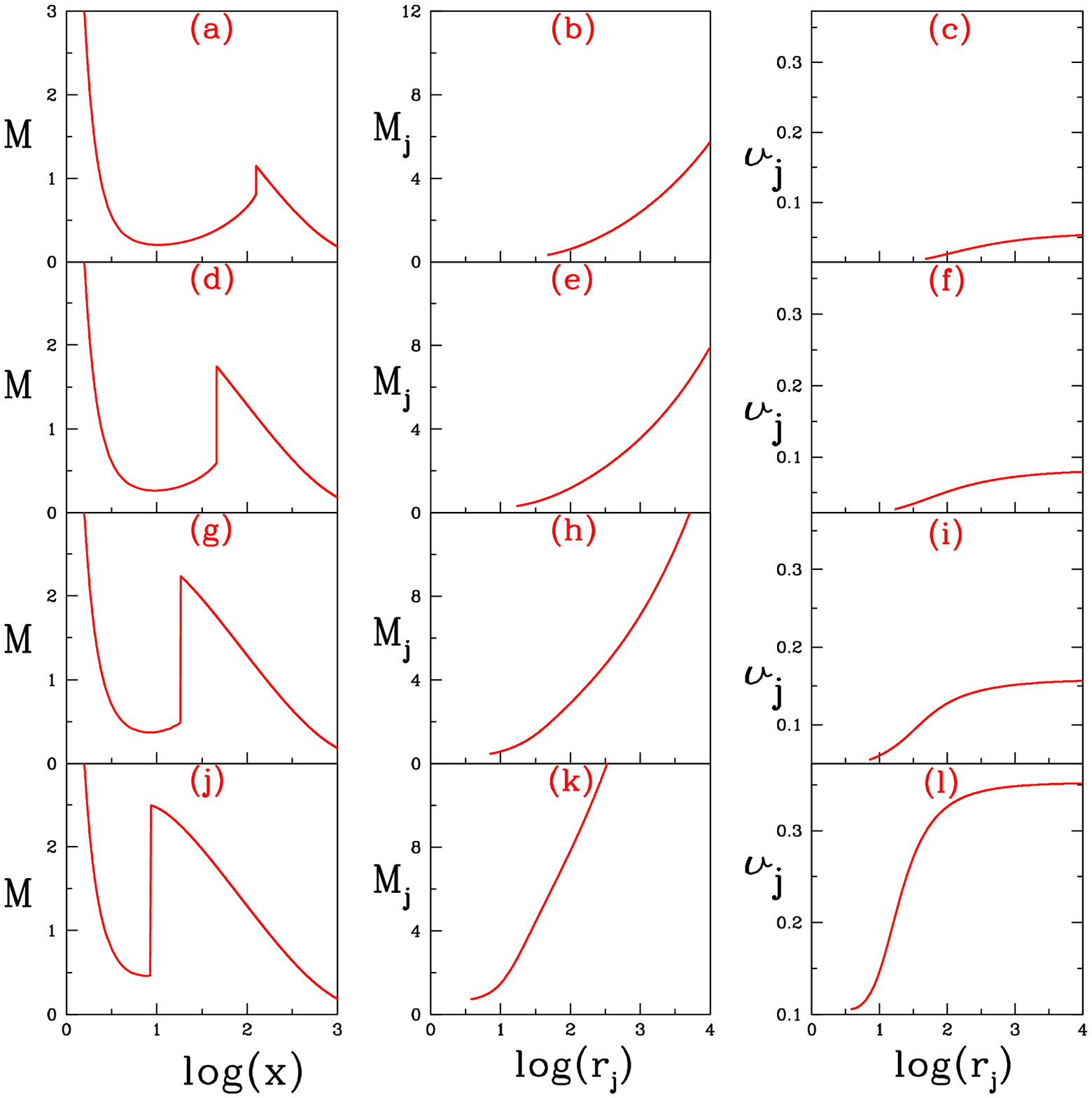,height=10.cm,width=10.cm,angle=0}
\caption{Variations of accretion Mach number $M$ (a, d, g, and j) with $log(x)$, $M_j$ (b, e, h, and k), and
$v_j$ (c, f, i, and l) with $log(r_j)$.The disc solutions are for parameters $E=0.001, \lambda_{inj}=18.592$ at
outer  boundary $x_{inj}=10^4$, and for $\alpha=0.009115$ (a-c), $0.009315$ (d-f), $0.009626$ (g-i), 
and $0.009875$ (j-l). Plots (a-c) are characterized by
$(x_s,~R_{\dot m},~\ell_{ps},~\ell_s=125.8964,0.0091,0.0054,0.0346)$;
for (d-f) $(x_s,~R_{\dot m},~\ell_{ps},~\ell_s=45.9986,0.0208,0.0103,0.0438)$; for (g-i)
$(x_s,~R_{\dot m},~\ell_{ps},~\ell_s=18.2871,0.0504,0.0380,0.0585)$; and for (j-l) 
$(x_s,~R_{\dot m},~\ell_{ps},~\ell_s=8.5741,0.0991,0.2337,0.0948)$,
and the jet terminal velocities are $v_{\infty}=0.0534,0.0793,0.1566$ and $0.3519$,
respectively.}
 \end{center}
\label{lab:fig9}
\end{figure}

 In Figs. 9a-l, we have plotted accretion
and jet solutions for various $\alpha$ and $\ell_{ps}$.
All the accretion solutions
($M$ with $log(x)$)
are for outer boundary parameters $E,~\lambda_{inj}~=~0.001,~18.592$ at the outer boundary $x_{inj}=10^4$. The
viscosity parameters are $\alpha=9.115\times10^{-3}$ (9a), $9.315\times 10^{-3}$ (9d), $9.626\times10^{-3}$ (9g), and
$9.875\times10^{-3}$ (9j). The vertical
jumps show the location of accretion shocks, and they are at $x_s=125.896$ (9a), $45.9986$ (9d), $18.2871$ (9g),
and $8.5741$ 9(j). The jet solutions corresponding to these accretion solutions, are presented by
$M_j$ (9b, 9e, 9h, 9k) and $v_j$ (9c, 9f, 9i, 9l).  
As $\alpha$ increases
$x_s$ decreases, therefore increasing the pre-shock disc. Moreover, with decreasing $x_s$, the post-shock disc becomes
smaller and hotter.
So as $x_s$ decreases, initially both $\ell_s$ \& $\ell_{ps}$ will increase, but at around $x_s\sim 100$, further
decrease of $x_s$ will reduce $\ell_s/\ell_{ps}$ and significantly
increase $\ell_{ps}$ (Fig. A1b). In Figs. 9a, d, g, j, increase of $\alpha$, causes a shift of $x_s = 125.896 \rightarrow 8.5741$.
Consequently, $\ell_{ps}$ increases from $0.0054 \rightarrow 0.2337$. 
The resulting jets are accelerated and the terminal velocity  increases from $v_{\infty}=0.0534
\rightarrow 0.3519$ as shock shifts from $x_s=125.896 \rightarrow 8.5741$, with the corresponding change in luminosity.
By considering the relative proportions of post-shock and pre-shock radiations, decrease of $x_s$ with increasing
$\alpha$ resembles the disc to move from hard state to hard intermediate state, and simultaneously the jet becomes
stronger and faster (both $v_{\infty}$ and $R_{\dot m}$ increases). In Fig. 10a-c, we show the comparison of shock
parameter space ($E-\lambda_0$) of the accretion disc without massloss (dotted, online red)
and with massloss but disc parameter
$\beta {\dot m}^2=0.001$ (long dashed, online black) and $\beta {\dot m}^2=0.01$
(dashed, online blue),
and for various viscosity parameter $\alpha=0$ (Fig. 10a), $\alpha=0.1$ (Fig. 10b) and $\alpha=0.2$ (Fig. 10c).
It is to be noted, that the bounded regions in $E-\lambda_0$ parameter space, show the parameters for steady
state shocks to occur, but non-steady shocks still exist outside the bounded region. The parameter space shrinks
when massloss is considered, because with massloss, the post-shock pressure decreases, and the entire range for which
steady shock may have existed in absence of jets, will not be able to satisfy the momentum balance across the shock
front. Moreover, shocks seem to exist for fairly high viscosity and in presence of massloss.

%%%%%%%%%%%%%%%%fig of ls/lps%%%%%%%%%%%%%%%%%%%%%%%%%

\begin{figure}
 \begin{center}
  \epsfig{figure=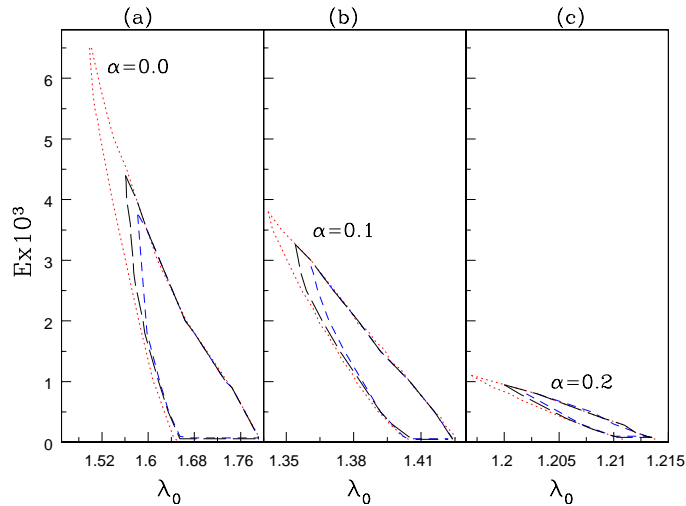,height=9.cm,width=9.cm,angle=0}
\caption{Shock parameter space or $E-\lambda_0$ space for accretion, considered without massloss (dotted),
and with massloss (dashed), and for $\alpha=0$ (a), $\alpha=0.1$ (b), and $\alpha=0.2$ (c).}
\end{center}
\label{lab:fig10}
\end{figure}

\section{Discussion and conclusion}

Our main focus in this paper has been to study radiative and thermal driving of bipolar jets from
dissipative accretion disc. The thermal driving is determined by the thermal energy
of the jet base or the post-shock region. In other words, hotter the post-shock region, faster
will be the jet. However, the thermal energy in the inner disc though hot, but cannot produce jets with $v_{\infty}>0.1$
\citep{c99,cd07,kc13}, for any realistic temperature ranges of the accretion disc. As has been explained in section 1,
only
thermally driven 
outflows can achieve terminal speeds of the order of the sound speed at the base.
Radiative driving of jet is
due to the momentum deposition of the accretion disc photons on to the shock driven jet. The issue of
radiative acceleration of jets have been studied earlier \citep{cc00,cc02}, however, the jet base conditions
in those works were assumed and were not self consistently obtained. In this paper, we do study the radiative acceleration of
jets, but also by self-consistently solving the jet solution from the accretion disc solution itself.
This is an exploratory study to investigate whether or not radiative driving is meaningful acceleration mechanism for jets.
Radiative moments were calculated a posteriori from viscous accretion solutions.

We initially considered the post-shock luminosity ($\ell_s$) as a parameter, and calculated the pre-shock
radiation from the disc solution ($a$, $\rho$) and by supplying $\beta {\dot m}^2$. 
Similar to \citet{kc13}, in this paper too, we find that with the increase of $\alpha$ for disc solutions starting
with same outer boundary condition, the shock location moves towards the horizon. 
As $x_s$ decreases for fixed values of $\ell_s$, the jets are stronger and faster
since the post-shock intensity increases (Figs. 6). For fixed values of
$\alpha$, if $\ell_s$ is increased, the jet is accelerated, as well as, more matter flows into the jet channel.
The depleted post-shock flow causes $x_s$ to decrease, in order to regain pressure balance across the shock.
Since enhanced $\ell_s$ increases the base velocity or $v_{jb}$, the jet sonic point decreases too.

The effect of pre-shock luminosity is quite different. Although pre-shock luminosity can accelerate as efficiently,
as the post-shock radiation, however, since pre-shock radiation can `see' mostly the supersonic branch
of the jet, so neither $R_{\dot m}$ nor $r_{jc}$ is affected appreciably. Therefore, one may conclude that post-shock
radiation both accelerates and controls the mass outflow rate, pre-shock radiation increases the kinetic energy of the jet.
And since for both pre-shock and post-shock radiations $2{\cal F}_{r_j} \sim ({\cal E}_{r_j}+{\cal P}_{r_j})$ at $r_j \sim$ few$\times
100r_g$, radiation drag effect is nullified at around that distance \citep{cdc04,c05}. Which means faster jets can be obtained if 
enough radiation power can be supplied. And indeed we obtained terminal speeds of $v_{\infty}\gsim 0.4$ for total disc
luminosity $\ell \sim 1$. Although we have used $\ell_s \leq 0.5$ (which is acceptable for luminous sources),
but assigning arbitrary values of $\ell_s$ may give wrong results, because $\ell_s$ crucially depends on pre-shock radiation.
So in Appendix 1, we have estimated $\ell_s/\ell_{ps}(x_s)$ from correct radiative losses from a set of solutions
of advective disc. We used this relation to estimate $\ell_s$, for given values of $\ell_{ps}$ and $x_s$. 
We then increased $\alpha$ for disc solutions starting with same $E~\&~\lambda_{inj}$, and calculated simultaneous
disc-jet solution. We showed that as $\alpha$ is increased, $x_s$ decreases and for reasons explained above,
not only the disc moved from low intensity disc to brighter disc, but the jet became stronger and faster.
Although, we have not considered the Keplerian disc component in our solution, but it resembles qualitatively,
the transition of the disc from hard to intermediate hard states with associated strengthening of the jet
from slow jets to faster and stronger jet, as has been reported in observations. We are still to include a few more physical
processes in our disc-jet model, but our results qualitatively shows the correlation of disc spectral states
and the jet states, as are observed in micro-quasars. Moreover, our result also indicates that, with multi-stage acceleration mechanism,
truly relativistic jets from accretion disc are a distinct possibility.

%%%%%%%%%%%%%%%%%%%SECTION 3 %%%%%%%%%%%%%%%%%%
%\section*{Acknowledgments}

%%%%%%%%%%%%%%%%%%%%%%%%%%%%%%%%%%%%%%%%%%%%%%%%%%%%%%%%

\appendix

%%%%%%%%%%%%%%%%%%%%%%%%%%%%%%%%%%%%%%%%%%%%%%%%%%%%%%%%%
\section{Estimation of post-shock luminosity from pre-shock radiations}
\label{app:spect}
We consider a general radiative transfer model (Chakrabarti \& Mandal 2006)
which consists of two components, a Keplerian disc on the equatorial plane and
a sub-Keplerian component on the top of the Keplerian disc. The Keplerian disc
supplies the multi-colour black body photons and a fraction of that photons are
inverse-Comptonized by post-shock region. The pre-shock sub-Keplerian disc
emits radiation via bremsstrahlung and synchrotron process whereas the
post-shock region produces the same as in pre-shock along with the
Comptonization of the local and intercepted soft photons. We calculate the radiation
spectrum of accretion disc using the viscous transonic shocked solution for a given
outer boundary condition $E=0.001, \lambda_{inj}=18.592$ at $x_{inj}=10^4$. For a given value of $\alpha$,
a shock is formed at $x_s$.
The frequency integrated pre-shock
and post-shock luminosity is then calculated.
It has been shown in \citep{kc13} that for the same set of boundary conditions,
$x_s$ decreases if $\alpha$ is increased. In our case for a range of
$0.0091 \leq \alpha \leq 0.01$, for flows starting with the same injection
values mentioned above, the shock location
changes from $x_s=131 \rightarrow 7.8$ (Fig. A1a). We calculate
the ratio of post-shock to pre-shock luminosity $\ell_s/\ell_{ps}$ for different
values of $x_s$ in the range mentioned above and it has been plotted by dots in
Fig. (A1b). In Fig. A1c we plot the associated variation in photon index,
which shows the spectra softens as shock moves inward.
The ratio ($\ell_s/\ell_{ps}$) increases initially as the shock
move inward, reaching a maximum value and then decreases sharply.
This behaviour is due
to fact that initially as shock moves inward both post-shock temperature and
density increases and hence post-shock luminosity increases but as shock reaches
closer to central object the pre-shock luminosity $\ell_{ps}$ increases
and the increased supply of pre-shock photons cools down the post-shock
flow, and therefore the ratio would decrease too.
We have fitted a polynomial through the model data points and this general
behaviour of $\ell_s/\ell_{ps}$ with $x_s$ is used to calculate
$\ell_s$ for given value of $\ell_{ps}$.

The fitted polynomial for the relation between post-shock and pre-shock disc
luminosity is,
\begin{equation}
f(x_s)=-0.659234+0.127851~x_s-0.00043~x^2_s-1.13\times10^{-6}~x_s^3,
\label{poly.eq}
\end{equation}
where, $f(x_s)$ is the fitted function for $\ell_s/\ell_{ps}$.

\begin{figure}
 \begin{center}
\epsfig{figure=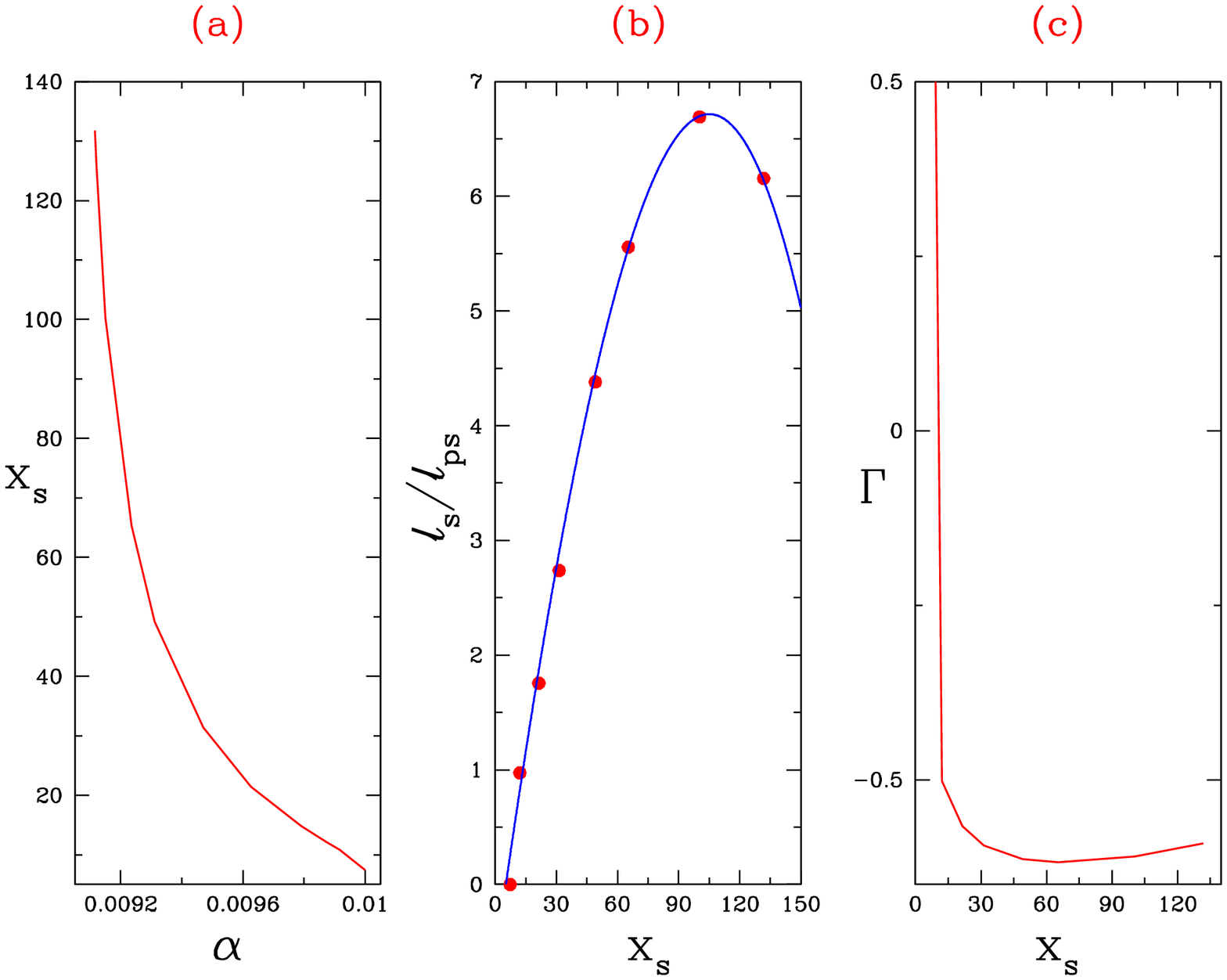,height=9.cm,width=9.cm,angle=0}
\caption{Variation of (a) shock location($x_s$) with disc viscosity parameter($\alpha$), and 
(b) ratio of $\ell_s/\ell_{ps}$ with shock locations, and (c) the photon index $\Gamma$
for the same injection values $E=0.001, \lambda_{inj}=18.592$ 
at disc outer boundary, $x_{inj}=10^4$ but varying viscosity parameter($\alpha$) from $0.0091$ to $0.01$. In 
plot (b) dot points are actual data points and solid line is the fitted polynomial.
 }
 \end{center}
\label{lab:figA2}
\end{figure}
 
 \label{lastpage}
%% %%\bibliographystyle{alpha} %% give your .bst file
%% %%\bibliography{ms_joshi}   %% give your .bib file

\end{document}